\documentstyle[12pt,aasms4]{article}
\received{: 1996 January 25}
\accepted{: 1996 October 10}
\lefthead{Corbel et al.}
\righthead{The distance of SGR 1806-20}

\begin{document}

\title{The Distance of the Soft Gamma Repeater SGR 1806-20}

\author{S. Corbel, \altaffilmark{1,2} P. Wallyn, \altaffilmark{3} T.M. Dame, \altaffilmark{4} P. Durouchoux, 
\altaffilmark{2}}
\author{W.A. Mahoney, \altaffilmark{3} O. Vilhu, \altaffilmark{5} and J.E. Grindlay \altaffilmark{4}}

%\and

% Notice that each of these authors has alternate affiliations, which
% are identified by the \altaffilmark after each name.  The actual alternate
% affiliation information is typeset in footnotes at the bottom of the
% first page, and the text itself is specified in \altaffiltext commands.
% There is a separate \altaffiltext for each alternate affiliation
% indicated above.

\altaffiltext{1}{Space Radiation Laboratory, 220-47, California Institute of Technology, Pasadena, CA 91125; corbel@aegir.caltech.edu. }
\altaffiltext{2}{DAPNIA, Service d'Astrophysique, CE Saclay, 91191 Gif sur Yvette Cedex, France.}
\altaffiltext{3}{Jet Propulsion Laboratory, 169-327, California Institute of Technology, 4800 Oak Grove Drive, Pasadena, CA 91109.} 
\altaffiltext{4}{Harvard-Smithsonian Center for Astrophysics, 60 Garden Street,
Cambridge, MA 02138.}
\altaffiltext{5}{Observatory, Box 14, FIN-00014 University of Helsinki,
Finland.}

% The abstract environment prints out the receipt and acceptance dates
% if they are relevant for the journal style.  For the aasms style, they
% will print out as horizontal rules for the editorial staff to type
% on, so long as the author does not include \received and \accepted
% commands.  This should not be done, since \received and \accepted dates
% are not known to the author.

\begin{abstract}
We present $^{12}$CO(J=1-0) observations in the direction of the Soft Gamma Repeater SGR 1806-20 with the SEST telescope.
We detected several molecular clouds, and we discuss in this paper the implications of these observations for the distance to the X-ray counterpart AX 1805.7-2025, the supernova remnant G10.0-0.3 and the very luminous O9-B2 star detected in the line of sight.
The distance of SGR 1806-20 is estimated to be 14.5 $\pm$ 1.4 kpc and this Soft Gamma Repeater is very likely associated with one of the brightest HII regions in the Galaxy, W31. 
The large size of G10.0-0.3 (25 $\times$ 38 pc) for a young supernova remnant possibly powered by a central pulsar (AX 1805.7-2025) indicates that G10.0-0.3 could be expanding in the very low density region produced by the wind of the blue star.
\end{abstract}

% The different journals have different requirements for keywords.  The
% keywords.apj file, found on aas.org in the pubs/aastex-misc directory, 
% contains a list of keywords used with the ApJ and Letters.  These are 
% usually assigned by the editor, but authors may include them in their 
% manuscripts if they wish. 


% That's it for the front matter.  On to the main body of the paper.
% We'll only put in tutorial remarks at the beginning of each section
% so you can see entire sections together.

\section{Introduction}

Soft Gamma Repeaters (SGRs) (Norris, Hertz \& Wood 1991) are believed to be a new class of high-energy transients that present recurrent outbursts with shorter duration and softer gamma ray spectra than Gamma-Ray Bursts (GRBs). 
Three SGRs have so far been detected, and they seem associated with supernova remnant (SNRs), although SGR 1806-20 is the only source whose position is known with high accuracy.
Indeed, one of its bursts has been simultaneously detected by the X-ray satellite {\em ASCA} ({\cite{mur94}) and the Burst and Transient Sources Experiment ({\em BATSE}) aboard the {\em Compton
Gamma Ray Observatory (CGRO)} (\cite{kou94}); its position coincides with those of the radio core of the supernova remnant G10.0-0.3 (\cite{kul94,vas95}) and a new quiescent X-ray source: AX 1805.7-2025 (\cite{mur94,son94}).
SGR 1900+14 (\cite{kou93}) the other SGR in the Galactic plane lies near the SNR G42.8+0.6 (\cite{vas94}), but we can not rule out an association with the superluminal source GRS 1915+105 (\cite{mir95}).
We note that the gamma-ray error box for SGR 0526-66 lies entirely within the SNR N49 in the Large Magellanic Cloud (\cite{cli82}), although the 1979 March event from this source was $\sim$10$^{4}$ times more luminous than the two SGRs in the Galactic plane. 
It is possible that SGRs may be associated with giant molecular clouds (GMCs) (\cite{gri94}).

Here we present the first accurate determination of the distance to a Soft Gamma Repeater: SGR 1806-20. The G10.0-0.3 SNR, surrounding SGR 1806-20, appears to be associated with the HII region W31, which is along the line of sight of this SGR. W31 is in turn associated with a large, well - defined molecular complex. 
We will present new evidence in support of the far kinematic distance for W31 and its associated giant molecular cloud and discuss the implications for SGR 1806-20.

We first report in \S\ 2 $^{12}$CO(J=1-0) observations toward SGR 1806-20 with the Swedish ESO Submillimeter Telescope (SEST) and then present new arguments in \S\ 3 to constrain the distance of W31.
A detailed analysis (\S\ 4) of the kinematics of the molecular clouds in the line of sight leads to a possible connection between SGR 1806-20, the SNR G10.0-0.3, W31, and a very bright blue star recently detected by Kulkarni et al. (1995).
Based on this analysis, we conclude that the distance to SGR 1806-20 is 14.5 $\pm$ 1.4 kpc; we discuss in \S\ 5 the consequences for the size and surrounding medium of the radio SNR and the luminosity of the blue star. Our conclusions are summarized in \S\ 6.

\section{Observations}

In 1995 April, we conducted millimeter observations with the 15 m SEST telescope situated on La Silla in Chile.
In order to study the distribution of the molecular gas along the line of sight, we observed the $^{12}$CO(J=1-0) emission at the position of the X-ray counterpart detected with {\em ASCA} at $\alpha$(1950)= $18^{h} 05^{m} 41^{s}$ and $\delta$(1950)= -20\arcdeg 25\arcmin 07\arcsec \ (\cite{mur94}). 
The half - power beamwidth (FWHM) of the telescope at 115 GHz is 45\arcsec. The front - end was a Schottky receiver with a typical system temperature (including atmosphere) of 700 K. 
The 1086 MHz bandwidth acousto - optical spectrometer (AOS) was used as the back - end, with a resolution of 1.4 MHz, giving a velocity resolution of 2.3 km s$^{-1}$. The intensity calibration was performed using the chopper-wheel method, and the results are given in terms of the main - beam brightness temperature T$_{mb}$ with an rms noise of 0.1 K (the main - beam efficiency of the SEST at 115 GHz is 0.7). 
Instrumental baseline structure was removed by position switching.

The $^{12}$CO(J=1-0) spectrum shown in Figure 1 is quite complex, with distinct molecular clouds at V$_{lsr}$= -16, 4, 13, 24, 38, 73, and 87 km s$^{-1}$ (hereafter MC -16, MC 4, MC 13, MC 24, MC 38, MC 73, and MC 87, respectively).
We also used the Galactic survey of Bitran (1987) to study the structure of the molecular clouds.
We discuss below the different arguments that constrain the distance to the molecular clouds along the line of sight and their positions compared to SGR 1806-20.

\section{The distance to the HII complex W31}

W31 is a large and complex HII region located in the Galactic plane around l=10$^{o}$.2 and b=-0.$^{o}$2. Three components of W31 present a particularly intense radio continuum: G10.2-0.3, G10.3-0.1, and G10.6-0.4. The G10.0-0.3 SNR is also clearly identified with some fainter emission (see Fig. 46 in Shaver \& Goss 1970). 
The closest source to the SNR G10.0-0.3, and by far the brightest in W31, is G10.2-0.3. Observations of the H110$\alpha$ recombination line by Downes al. (1980) toward G10.2-0.3 showed a peak at a velocity of 13\ km\ s$^{-1}$.
For a more complete review of W31 see Ghosh et al.\ (1989).

In Figure 2a, we present a map of the molecular cloud at 13 km s$^{-1}$ (MC 13) superimposed on the 4.8 GHz radio continuum contour of G10.2-0.3 and G10.3-0.1.
The $^{12}$CO(J=1-0) map of MC 13 and the radio continuum contours of W31 seem to be in good agreement with each other.
In light of the very close coincidence of MC 13 and W31, both spatially and in velocity, and of the close association of HII regions with molecular clouds in general, it is very likely that W31 and MC 13 are associated.

With this recombination line at 13 km s$^{-1}$, and using the Galactic rotation curve of Burton (1988), we deduce a kinematic distance of 2.3 or 14.5 kpc for W31.
Moreover, H$_{2}$CO, OH and HI absorption lines against the HII region continuum are detected up to 46 km s$^{-1}$. Therefore, the HII complex has to be situated behind the clouds detected at the position of the absorption lines. Thus we can rule out the near distance. 
Based on the lack of HI absorption lines greater than 46 km s$^{-1}$, it has been argued by Wilson (1974) and Kalberla, Goss \& Wilson (1982) that W31 is not at the far kinematic distance of 14.5 kpc either, but rather in or near the expanding 3 kpc Arm at a distance of $\sim$5 kpc. This arm has a large proper motion (expansion velocity of $\sim$30 km s$^{-1}$) compared with the expected V$_{lsr}$ from purely circular Galactic rotation. This became the ``standard'' distance for W31.

However, absorption at velocities greater than 46 km s$^{-1}$ is {\it unlikely}, and therefore there is no reason to reject the far distance. 
Indeed, at low Galactic longitudes (5$^{o}$ $\sim$ 20$^{o}$) absorption by H$_{2}$CO and cold HI should in general be very rare at velocities greater than 46 km s$^{-1}$ because there is very little gas in any form at such velocities (see, e.g., Fig. 3 in \cite{dam87}). 
This ``hole'' in the gas distribution at R $<$ 3 kpc is a well-known feature of the Galaxy (e.g., \cite{bur76}) that was not discussed by Kalberla et al.\ (1982). 
One might expect that the cloud we detected at 73  km s$^{-1}$ (this single isolated cloud is visible in  the CO longitude-velocity map shown in Fig. 3) would be seen in absorption, but as Figure 2b shows, the line of sight to G10.2-0.3 (by far the brightest component of W31) passes just {\it below} the edge of this cloud.
The absence of significant CO emission near 73 km s$^{-1}$ toward G10.2-0.3 is confirmed by the higher angular resolution survey of Sanders et al. (1986) - see their b-v map at l = 10.$^{o}$2.
Therefore, the lack of absorption against G10.2-0.3 observed by Kalberla et al. (1982) near 73 km s$^{-1}$ is not surprising.
It is worth noting that Greisen \& Lockman (1979) detected, in a previous observation in the same direction, a weak absorption feature at 75 km s$^{-1}$ in agreement with the position of MC 73 and therefore adopted the far distance for W31.
The position we observed (marked ``SGR 1806-20'' in Fig. 2b) is significantly closer to the center of this cloud, and therefore it is not surprising that we detected it in CO emission.
An H$_{2}$CO absorption line measurement at the exact angular position of AX 1805.7-2025 could  be very useful to confirm the association of W31 with
MC 13, the position of MC 73 relative to W31, and the distance to W31.

Although there is little reason to reject the far distance of 14.5 kpc for MC 13\ - and in fact we will present evidence below that {\it favors} it\ - the anomalous velocity of $\sim$30 km s$^{-1}$ implied for MC 13 by situating W31 at 5 kpc is very strong evidence {\it against} it being at that distance.
Indeed, it is well established that molecular clouds are a very ``cold'' Galactic population with a cloud-cloud velocity dispersion of $\sim$4.2 km s$^{-1}$ (\cite{com91}). 
Thus, the only plausible way that a large molecular cloud such as MC 13 ($\sim$4 $\times$ 10$^{5}$ M$_{\sun}$ at a distance of 5 kpc) could have an anomalous velocity of 30 km s$^{-1}$ is if it is associated with the expanding 3 kpc Arm, and that is indeed what Kalberla et al.\ (1982) and Wilson (1974) suggested.

However it is very unlikely that MC 13 is associated with the 3 kpc Arm, since:
(1), as Figure 3 shows, the 3 kpc Arm is a well-defined linear feature in the CO longitude-velocity diagram with a velocity width of only $\sim$10 km s$^{-1}$ (and thus a velocity dispersion about the mean, $\sigma_{V} \le$ 5 km s$^{-1}$).
At $l$ = 10$^{o}$ the arm is centered near -10 km s$^{-1}$, and so MC 13 lies $\sim$23 km s$^{-1}$ ($\sim$5 $\sigma_{V}$) outside the central velocity of the arm.
(2) The 3 kpc Arm is known to be quite deficient in star formation (\cite{loc89}), while W31 is one of the brightest HII regions in the entire Galaxy.
(3) As Figure 3 shows, MC 13 is far brighter in CO (and thus, far more massive) than any cloud in the 3 kpc Arm.

Additional evidence that MC 13 lies at the far distance of 14.5 kpc comes from comparing its angular size with its line width. It has been well established by many studies that molecular clouds of all sizes exhibit an empirical relation between size and line width (e.g., \cite{dam86}).
It is obvious even from Figure 3 that MC 13 has a small angular size and a large velocity width related to other clouds nearby in the l-v map. 
The large line width implies that the cloud is massive, and the small angular size implies that it is far away. With line width and angular size defined as in Dame et al.\ (1986), the far distance is clearly favored by the radius - line width relation ($\Delta$V=16 km s$^{-1}$, R=63 pc for a distance of 14.5 kpc).
It is worth mentioning that the active star formation indicated by the presence of the very bright W31 HII region and the SNR might be expected to increase the line width of MC 13 somewhat above the normal R - $\Delta$V relation, as observed.

MC 13 is at a sufficiently large Galactic radius ($\sim$6 kpc) that the Galactic rotation is quite orderly and well measured (\cite{bur88}). 
Probably the most significant source of error on the kinematic distance of MC 13 is therefore the cloud-cloud velocity dispersion of molecular clouds (4.2 km s$^{-1}$; \cite{com91}). 
If we conservatively fold in an additional uncertainty of 10 km s$^{-1}$ due to possible systematic non circular motions, such as those associated with spiral arms, we obtain a 1 $\sigma$ error of 1.4 kpc on the far kinematic distance of MC 13. 
It should be emphasized that uncertainties related to our extinction estimate in \S\ 4.2 (e.g., on the CO - to - H$_{2}$ conversion factor, the gas-to-dust ratio, HI optical depth, etc.) do not have any direct influence on our distance uncertainty, since the distance is determined purely from the cloud's systemic velocity, and this is known to an accuracy of a few kilometers per second.  
We can therefore conclude that W31 is at the distance of 14.5 $\pm$ 1.4 kpc.

\section{The distance to SGR 1806-20}

\subsection{Position of the Different Molecular Clouds}

We will now discuss the distances to the detected molecular clouds and their positions relative to SGR 1806-20 using our $^{12}$CO(J=1-0) spectrum toward AX 1805.7-2025 (Fig. 1). The parameters of the clouds are presented in Table 1.
%\placetable{table1}
In order to study the sizes of the clouds, we used the CO survey of Bitran (1987).
The differential Galactic rotation curve (\cite{bur88}) implies two possible distances for each of the molecular clouds at positive velocity, and one for MC -16. As we showed in \S\ 3, MC 13 is associated with W31 and therefore is at 14.5 kpc.
Since W31 and MC 13 are at the far distance, all the clouds at higher velocity (i.e., all except MC 4 and MC -16) must lie in front of them.
In the survey of Downes et al.\ (1980), H$_{2}$CO absorption lines against the G10.2-0.3 radio continuum were detected at V$_{lsr}$ = 1.4, 8.5, 16.6, 26, 32, and 36 km s$^{-1}$ (line widths of 6, 7, 4, 3, 3, and 2 km s$^{-1}$, respectively). The lines at 1.4, 26, and 36 km s$^{-1}$ are compatible with the mean V$_{lsr}$ of the molecular clouds we detected at 4, 24, and 38 km s$^{-1}$ if we take into account the velocity resolution of the AOS and the measured width of the CO lines.
These measurements confirm that MC 24 and MC 38 are in front of the W31 complex and imply that MC 4 is located at the near distance of 1 kpc.
We have already discussed in \S\ 3 the lack of absorption near 73 km s$^{-1}$. There is no need to resolve the distance ambiguity for MC 24, MC 38, MC 73, and MC 87, since they are in front of W31 at either distance.

The kinematic distance of the molecular cloud at -16 km s$^{-1}$ (MC -16, FWHM=10 km s$^{-1}$) is $\sim$22.5 kpc, far outside the orbit of the Sun around the Galactic center. 
However, an HI absorption feature at V$_{lsr}$ = -20 km s$^{-1}$ found in the direction of G10.3-0.1 and G10.2-0.3 (Greisen \& Lockman 1979; \cite{kal82}) implies that MC -16 is also in front of W31. 
To explain both the negative V$_{lsr}$ and a closer distance, MC -16 must have a non circular motion. As Figure 3 shows, MC -16 lies very close to the 3 kpc Arm. In \S\ 3, we showed that at this Galactic longitude the arm is centered near -10 km s$^{-1}$, very close to the velocity of MC -16.
It is therefore very likely that MC -16 is located in the 3 kpc Arm because (1) an absorption line is detected at -20  km s$^{-1}$, (2) MC -16 is nearly coincident in velocity with the 3 kpc Arm (1 $\sigma$ in velocity from the 3 kpc Arm), (3) the  angular size of MC -16 implies a diameter of 280 pc at a distance of 22.5 kpc but a much more reasonable diameter of 62 pc at 5 kpc, and (4) molecular clouds beyond the solar orbit are rare.
Although the R - $\Delta$V relation ($\Delta$V=18 km s$^{-1}$, R=31 pc for a distance of 5 kpc) for normal disk clouds is in favor of the far distance; clouds in the 3 kpc Arm tend to have unusually large line widths (\cite{dam96}), which may be related to the possibly explosive origin of the 3 kpc Arm or the presence of a bar in the center of our Galaxy.
Therefore, MC -16 probably has a proper motion relative to the expected V$_{lsr}$ from a pure Galactic rotation speed and a distance of $\sim$5 kpc is thus deduced. In Figure 4 we present a schematic diagram of the positions of the different molecular clouds.

\subsection{SGR 1806-20}	

We now have a better idea of the position of the molecular clouds: MC 13 is associated with W31 at 14.5 kpc, and all the other clouds are closer.
To calculate the total H$_{2}$ column density, N(H$_{2}$), due to the molecular clouds detected in the line of sight, we use our $^{12}$CO spectrum (Fig. 1) and we use the standard estimate of the ratio (Solomon \& Barrett 1991):
\begin{equation}
\frac{N(H_{2})}{\int T_{mb} (^{12} CO) dv } = 2 \times 10^{20} \mathrm{ \ molecules \ cm^{-2}/(K \ km \ s^{-1})}
\end{equation}
The uncertainty on N(H$_{2}$) inferred from this relation is approximately a factor of 2 (W. Langer, 1995, private communication). The $^{12}$CO(J=1-0) transition is the only line that has been properly calibrated as an H$_{2}$ mass tracer on a Galactic scale. For estimating H$_{2}$ column densities across the Galaxy, $^{12}$CO is preferable to $^{13}$CO (\cite{com91}).
After doing the $^{12}$CO\ -\ N(H$_{2}$) conversion, it is possible to derive the associated visual extinction, A$_{V}$, from the ratio $\frac{N(HI) + 2N(H_{2})}{E(B-V)} = 5.8\ \times \ 10^{21}$ atoms cm$^{-2}$ mag$^{-1}$ (Bohlin, Sauvage \& Drake 1978) and using $\frac{A_{V}}{E(B-V)} = 3.1 $, also valid at high extinction (\cite{dic78}). In a molecular cloud the hydrogen is molecular in form, so $\frac{N(H_{2})}{A_{V}} = 0.9 \times 10^{21}$ molecules cm$^{-2}$ mag$^{-1}$. All our results are summarized in Table 1.

Kulkarni et al.\ (1995) performed visible and near infrared observations in the direction of SGR 1806-20 and revealed the presence of several stars in the X-ray error box of AX 1805.7-2025.
Among them is a bright blue star (exactly coincident with the core of the G10.0-0.3 SNR), possibly a Luminous Blue Variable (LBV), with a high visual extinction (A$_{V} \sim$30) in agreement with the {\em ASCA} X-ray absorption measurement (\cite{son94}). Subsequently, van Kerkwijk et al.\ (1995) obtained near infrared spectra of this source and detected a He absorption line from which they deduced a spectral type of O9-B2. 
Even assuming extinction from only MC 73, a molecular cloud quoted in Grindlay\ (1994) at a distance of 6 kpc, they deduced that this star had to be one of the brightest in our Galaxy with a bolometric luminosity greater than 10$^{6}$ L$_{\sun}$. 

An X-ray spectrum of AX 1805.7-2025 by {\em ASCA} gives an absorbing neutral hydrogen column density N(H) = (6.0 $\pm$ 0.2) $\times$ 10$^{22}$ cm$^{-2}$ for a single power law fit and  N(H) = (5.2 $\pm$ 0.2) $\times$ 10$^{22}$ cm$^{-2}$ for a thermal Bremsstrahlung model from which a visual extinction of A$_{V} \sim$30 is also deduced (\cite{son94}).
 
If we integrate all H$_{2}$ material deduced from CO emissions in front of W31 (this means all the molecular clouds except MC 13) we calculate an extinction of 31 $\pm$ 5 mag. 
We must also consider dust associated with atomic hydrogen for the origin of the extinction.  The 21 cm spectrum in Figure 1 is from the Leiden-Dwingeloo survey of neutral hydrogen (\cite{har94}; Hartmann \& Burton 1996).
All the HI emission from V$_{lsr}$ = 13 km  s$^{-1}$ to infinity can be assumed to arise between us and MC 13, as well as about half the emission in the range 0 to 13 km  s$^{-1}$.
If we make the usual assumption that the HI emission is optically thin, one can deduce the HI column density from the brightness temperature (\cite{bur88}): 
\begin{equation}
N(HI) = 1.8\ \times \ 10^{18} \int T_{b} dv \mathrm{ \ atoms \ cm^{-2} }
\end{equation}
We estimate 4.5 mag as the contribution from neutral hydrogen to the total extinction in front of W31.
Therefore, we achieve a total visual extinction of 35 $\pm$ 5 mag along the line of sight without including MC 13. If we add the contribution of MC 13 (A$_{V}$ = 11 $\pm$ 3 mag.), we then have a total extinction of 46 $\pm$ 6 mag.

The high intrinsic luminosity of the blue star makes it one of the natural components of the origin of the bright W31 complex and the massive GMC MC 13 a probable progenitor. Moreover, the total visual extinction favors a position of this O9-B2 star near the front edge of MC 13.
AX 1805.7-2025 shows the same extinction as the previous star, and seems to be the core of the G10.0-0.3 SNR, which is therefore also located in W31.
It is even possible that part of MC 13 also contributes to the extinction  of SGR 1806-20. 
There is an H$_{2}$CO absorption line at 16.6 km s$^{-1}$ ($\Delta$V=4 km  s$^{-1}$, \cite{dow80}) detected against the radio continuum of G10.2-0.3, which is compatible with the velocity of MC 13. In Figure 2a the SNR G10.0-0.3 seems to be located on the edge of MC 13 (the position of SGR 1806-20 is the same as the core of G10.0-0.3). 

A likely scenario is that W31 is inside MC 13. This massive cloud gives birth to massive stars that create the W31 HII region, and some of these stars burn all their material and become SNRs. 
W31 would then be embedded inside MC 13 and extend to its edge, where the G10.0-0.3 SNR is located. The star formation continues inside MC 13. 

From (1) the spatial coincidence of AX 1805.7-2025 and the LBV star with SGR 1806-20, (2) the probable association of the SNR and the LBV star with W31 and MC 13, (3) the presence of a LBV star with the same absorption as the X-ray counterpart of SGR 1806-20, and (4) the measured absorption of all the material along the line of sight, it is possible to deduce that both SGR 1806-20 and the O9-B2 star probably reside in W31, i.e., on the edge of the giant molecular cloud at V$_{lsr}$ = 13 km  s$^{-1}$  and thus at the distance of 14.5 $\pm$ 1.4 kpc. 
Let us nevertheless point out that our conclusion is based on the assumption that there is no self - absorption from that source. We discuss below the possible connection between SGR 1806-20 and the O9-B2 star.

\section{Discussion}

\subsection{The Radio Nebula G10.0-0.3}

The radio  nebula G10.0-0.3 consists of nested shells and a bright central peak. Kulkarni et al.\ (1994) suggested that it consists of a young isolated pulsar powering a surrounding 9 $\times$ 6 arcmin$^{2}$ SNR with relativistic particles. 
From our distance evaluation, we thus deduce that the SNR has a size of 25 $\times$ 38 pc$^{2}$.

Kafatos et al.\ (1980) studied the interaction between supernova ejecta and their environments. In the rarefied interstellar medium (ISM), after the initial supernova explosion, the ejecta expand freely without bound (free - expansion phase) and then encounter significant gas to form a shock and a shell, that is detectable as a radio SNR.
For a SNR expanding in the ISM, Wheeler et al.\ (1980) find a typical SNR radius of $\sim$3.1 pc at the end of the free - expansion phase and 0.12 pc for a SNR expanding in a molecular cloud with density of n $\sim$ 10$^{4}$ cm$^{-3}$.
Kulkarni et al. (1994) favor a young pulsar of age less than 10$^{4}$ yr at the origin of the steady X-ray source AX 1805.7-2025. Using Table 1 in Kafatos et al.\ (1980), we found that the size of the SNR of 25 $\times$ 38 pc$^{2}$ (radius of $\sim$16 pc) and the approximate age of the pulsar favor a third possibility 
that implies a possible connection between the O9-B2 star and the SNR: the expansion of the G10.0-0.3 SNR in the hot and highly evacuated volume produced by the wind from the bright O9-B2 star. 
Indeed, Kafatos et al.\ (1980) found typical radii of the order of $\sim$19 pc at the end of the free expansion phase (age $<$ 3.7 $\times$ 10$^{3}$ yr) for a SNR expanding in a very low density region.

\subsection{The Luminous Star in the Direction of SGR 1806-20}

With our distance evaluation, the O9-B2 star is very luminous with nearly $\sim$ 5 $\times$ 10$^{6}$ L$_{\sun}$, and is perhaps the brightest star in the Galaxy. It is of course still not
clear what is the physical connection of this LBV star and the SGR source, although the
extreme luminosity suggests an association. We note that several stars not resolved by the telescope could produce the same result. High resolution infrared observations of this star should be very useful to solve the problem. If the ``LBV star'' is actually a cluster, this might also be
instead hot gas and wind heated by accretion onto a single massive object in the GMC, as
suggested in Grindlay (1994) for SGR sources in general.

\section{Conclusions}

By studying the kinematics of the molecular clouds along the line of sight to W31, we draw the conclusion that this HII complex is located at the far kinematic distance of 14.5 $\pm$ 1.4 kpc. This HII region is probably created inside the giant molecular cloud detected at 13 km s$^{-1}$ (MC 13).
At the far distance, MC 13 is one of the most massive molecular complexes in the Galaxy ($\sim$4\ $\times$ 10$^{6}$ M$_{\sun}$).
It is not surprising that such a cloud would  be associated with one of the Galaxy's brightest HII regions, and one of its brightest stars (or compact clusters).

Based on the apparent association of SGR 1806-20, the SNR G10.0-0.3, the HII region W31, and the GMC at 13 km s$^{-1}$, it was possible to constrain the distance to SGR 1806-20 to 14.5 $\pm$ 1.4 kpc with some confidence. The derived extinction deduced from CO and HI observations is consistent with that distance. It is even possible that the G10.0-0.3 SNR expands into the wind swept region from the O9-B2 star. 

Scheduled millimeter observations of molecules with enhanced abundances in high-ionization environments such as HCO$^{+}$, HCN, and CN (\cite{dur96}) or the study of shock tracers such as the IR transition at 1.58 \micron \ of C$^{+}$ (S. Kulkarni, 1995, private communication) could be very useful to confirm the association of the SNR with MC 13.
Continuum measurements at $\lambda$ = 1-3 mm of the thermal emission from dust and IRAS high resolution maps have been published elsewhere (\cite{wal95,smi96,van96}). Our results agree with the
scenario of a young neutron star in a SNR for SGR 1806-20 but the mechanism at the origin of the short X-ray bursts detected remains unclear.
The accurate distance evaluation of SGR 1806-20 could yield new constraints on SGR emission theories (e.g., \cite{ulm94,bar95}; Melia \& Fatuzzo 1995; Thompson \& Duncan 1995, Duncan \& Thompson 1996).

Note added in manuscript: We recently observed the 23.694 GHz line of the NH$_{3}$ transition against the radio continuum of the core of the G10.0-0.3 SNR and we detected an absorption line at a velocity of 73 km s$^{-1}$.
This detection confirms that the SNR and its associated molecular cloud MC 13 lie behind MC 73, and therefore must be at the far kinematic distance, as we proposed.

\acknowledgments

The authors would like to thank D. Hartman and W. Burton for providing the HI data before publication. S. C. and P. W. thank R. Duncan, M. G\'erin, S. Kulkarni, W. Langer, T. Murakami, T. Phillips, N. Scoville, and I. Smith for helpful and stimulating discussions.
P. D. and O. V. thank F. Azagra for his competent support during the observation with SEST. S. C. is also grateful to T. Hunter and D. Williams for their help. This work was supported in part by a grant from the National Science Foundation.
The Swedish-ESO Submillimeter Telescope is operated by the Swedish National Facility for Radio Astronomy, Onsala Space Observatory at Chalmers University of Technology, and by ESO.

\clearpage
\figcaption[figure1.eps]{$^{12}$CO(J=1-0) spectrum in the direction of SGR 1806-20. The dotted line represents the HI spectrum at l=10$^{o}$ and b=-0.$^{o}$25 (intensity divided by 14), an average of the spectra at (10,0) and (10,-0.5) from the Leiden-Dwingeloo survey of neutral hydrogen (Hartmann 1994; Hartmann \& Burton (1996)).\label{fig1}}

\figcaption[SGR_fig2.eps]{(a)$^{12}$CO(J=1-0) map of the molecular cloud detected at 13 km s$^{-1}$. The emission is integrated over the range 2 to 20 km s$^{-1}$. The contours at 4850 MHz of G10.2-0.3 and G10.3-0.1 are over plotted (contours at 0.1, 1, 2, 4, 8, 16, 32, 64 Jy beam$^{-1}$). The dot indicates the position
of SGR 1806-20.
\protect\newline
(b) Same as (a), but for the molecular cloud at 73 km s$^{-1}$. The CO emission is integrated from 60 to 80 km s$^{-1}$. These maps are derived from Bitran (1987).\label{fig2}}

\figcaption[SGR_fig3.eps]{Longitude - velocity map of Galactic CO emission integrated in the range from -0$\fdg$5 to 0$\fdg$5 in latitude from the survey of Bitran (1987). The contour interval is 0.5 K deg \label{fig3}}
 
\figcaption[figure4.eps]{Schematic diagram of the molecular clouds in the line of sight of SGR
1806-20. The clouds MC 24, MC 38, MC 73 and MC 87 were arbitrarily assigned to their near kinematics distances, but at either distance they lie closer than MC 13. The A$_{V}$ represent the visual extinction of the molecular clouds. \label{fig4}}
 
\clearpage

\begin{deluxetable}{cccccccc}
%\footnotesize
\scriptsize
\tablecaption{Derived Parameters from the $^{12}$CO(1-0) Spectrum toward SGR 1806-20}
\label{table1}
\tablecomments{We did not attempt to resolve the distance ambiguity for MC 24, MC 38, MC 73 and MC 87. A$_{V}$ represents the visual extinction of the cloud, and $\Delta$A$_{V}$ its uncertainty.}
\tablewidth{0pc}
\tablehead{
\colhead{V$_{lsr}$} & \colhead{$\Delta$V(FWHM)} & \colhead{Near Distance} &
\colhead{Far Distance} & \colhead{Estimated Distance} & 
\colhead{N(H$_{2}$)} & \colhead{A$_{V}$} & \colhead{$\Delta$A$_{V}$} \\
\colhead{(km s$^{-1}$)} &  \colhead{(km s$^{-1}$)} & \colhead{(kpc)} &
\colhead{(kpc)} & \colhead{(kpc)} & \colhead{(10$^{21}$ cm$^{-2}$)} & 
\colhead{(mag.)} & \colhead{(mag.)} 
}
\startdata
-16 & 10 & N. A. & 22.5 & 5.0 & 8.8 & 9 & 2\nl
4 & 5 & 1.0 & 16.0 & 1.0 & 3.7 & 4 & 1\nl
13 & 7 & 2.3 & 14.5 & 14.5 & 10.5 & 11	& 3\nl
24 & 5 & 3.2 & 13.5 & ? & 6.2 & 7 & 2\nl
38 & 7 & 4.6 & 12.3 & ? & 2.9 & 3 & 1\nl
73 & 10 & 6.2 & 10.5 & ? & 6.8 & 7 & 2\nl
87 & 6 & 6.8 & 10.2 & ? & 1.4 & 1 & 0.5\nl
\enddata
\end{deluxetable}
\clearpage

%\begin{deluxetable}{lccc}
%\footnotesize
%\tablecaption{Parameters used for the R-$\Delta$V relation}
%\label{table2}
%\tablecomments{R$_{1}$ defines the radius of the cloud if it is at the near distance and R$_{2}$ for the far distance. $\Delta$V is calculated from an average spectrum of the cloud.}
%\tablewidth{0pc}
%\tablehead{
%\colhead{  } & \colhead{$\Delta$V(FWHM)} & \colhead{R$_{1}$} & \colhead{R$_{2}$} \\
%\colhead{  } & \colhead{(km s$^{-1}$)} & \colhead{(pc)} & \colhead{(pc)} 
%}
%\startdata
%MC -16 & 18 & 31 & 141 \nl
%MC 13  & 16 & 22 & 63  \nl
%MC 73 & 14 & 23 & 38  \nl
%\enddata
%\end{deluxetable}

\clearpage

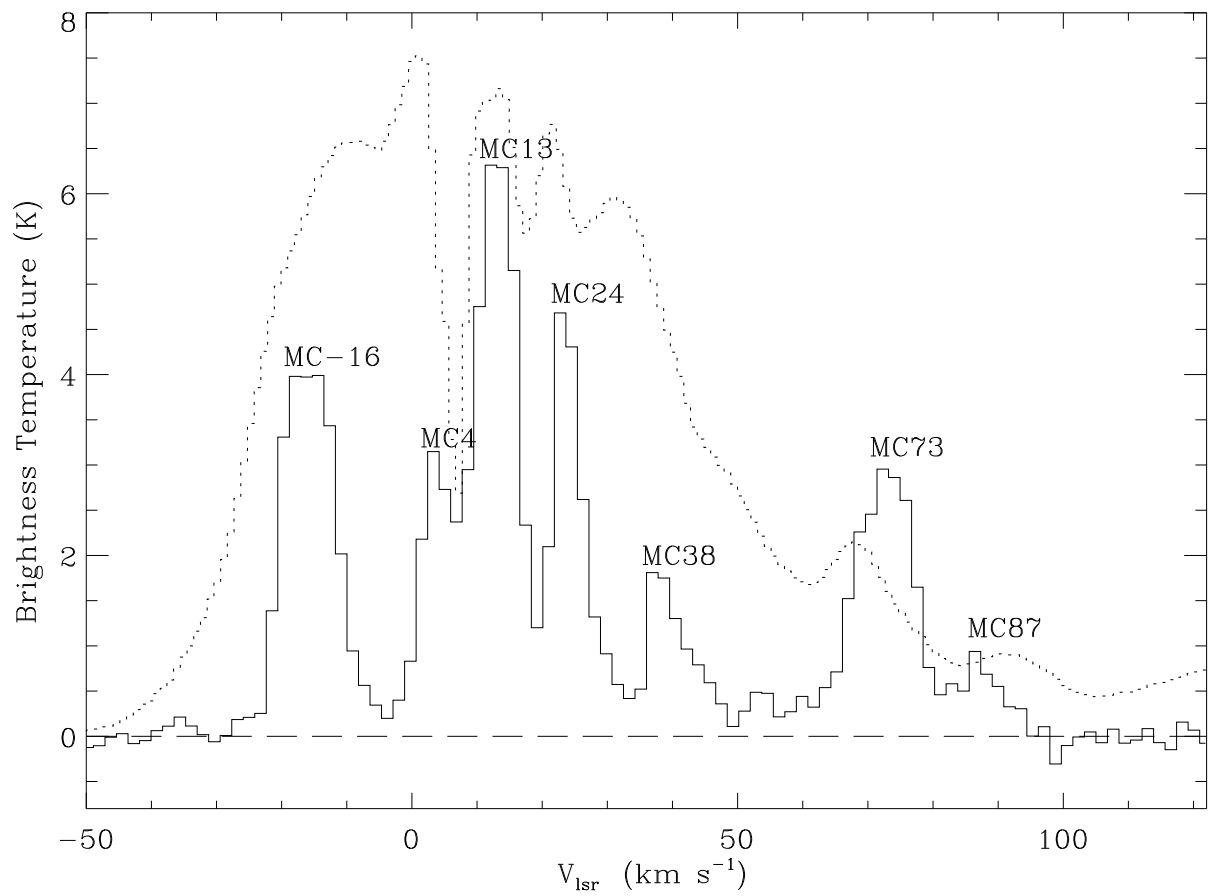
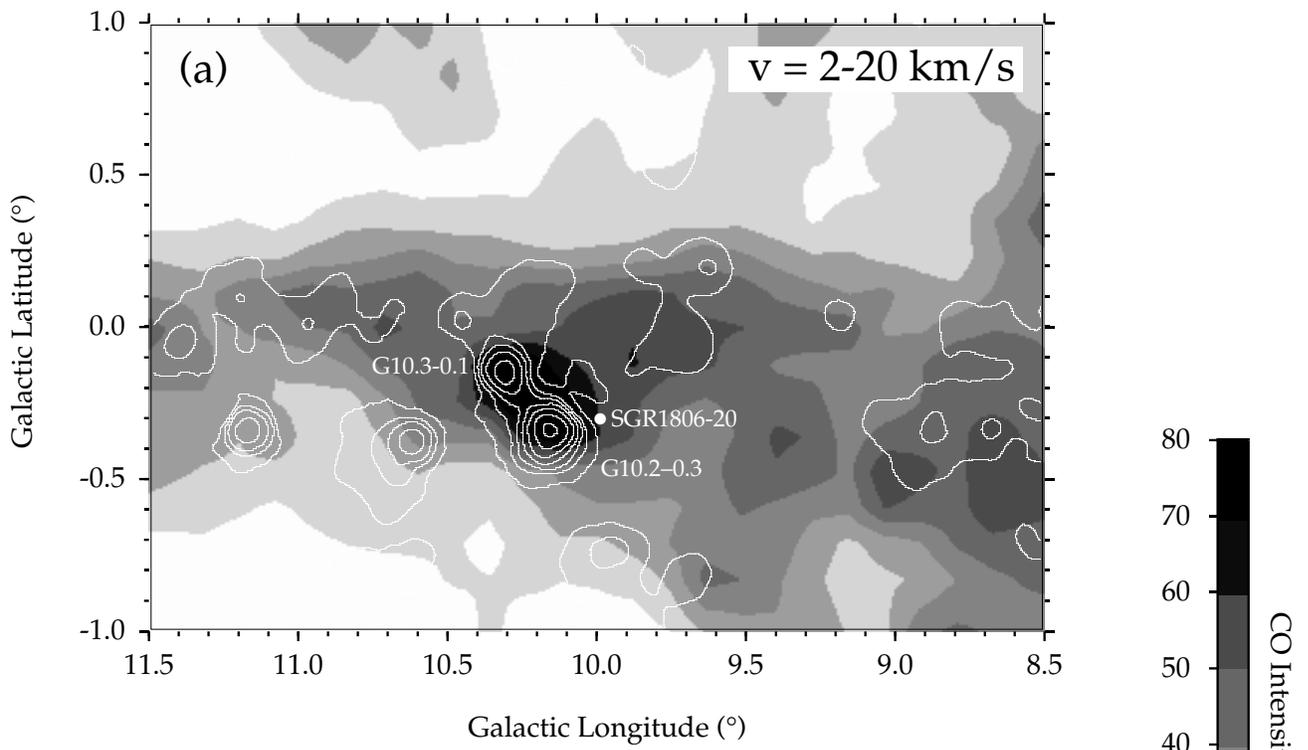
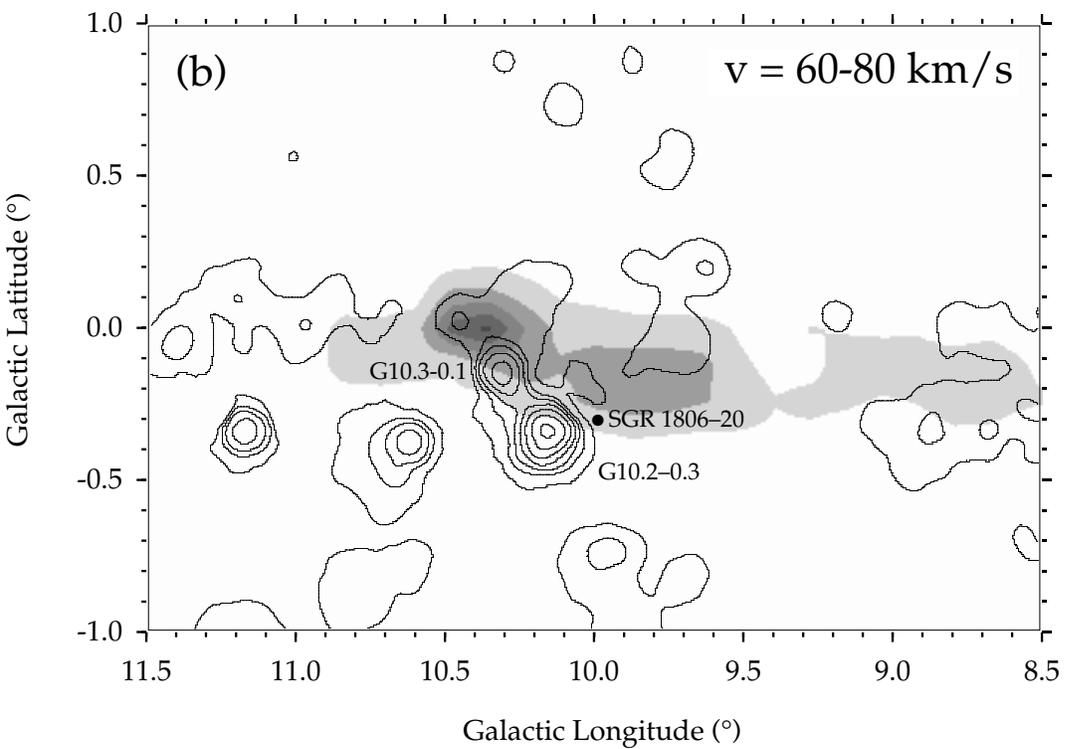
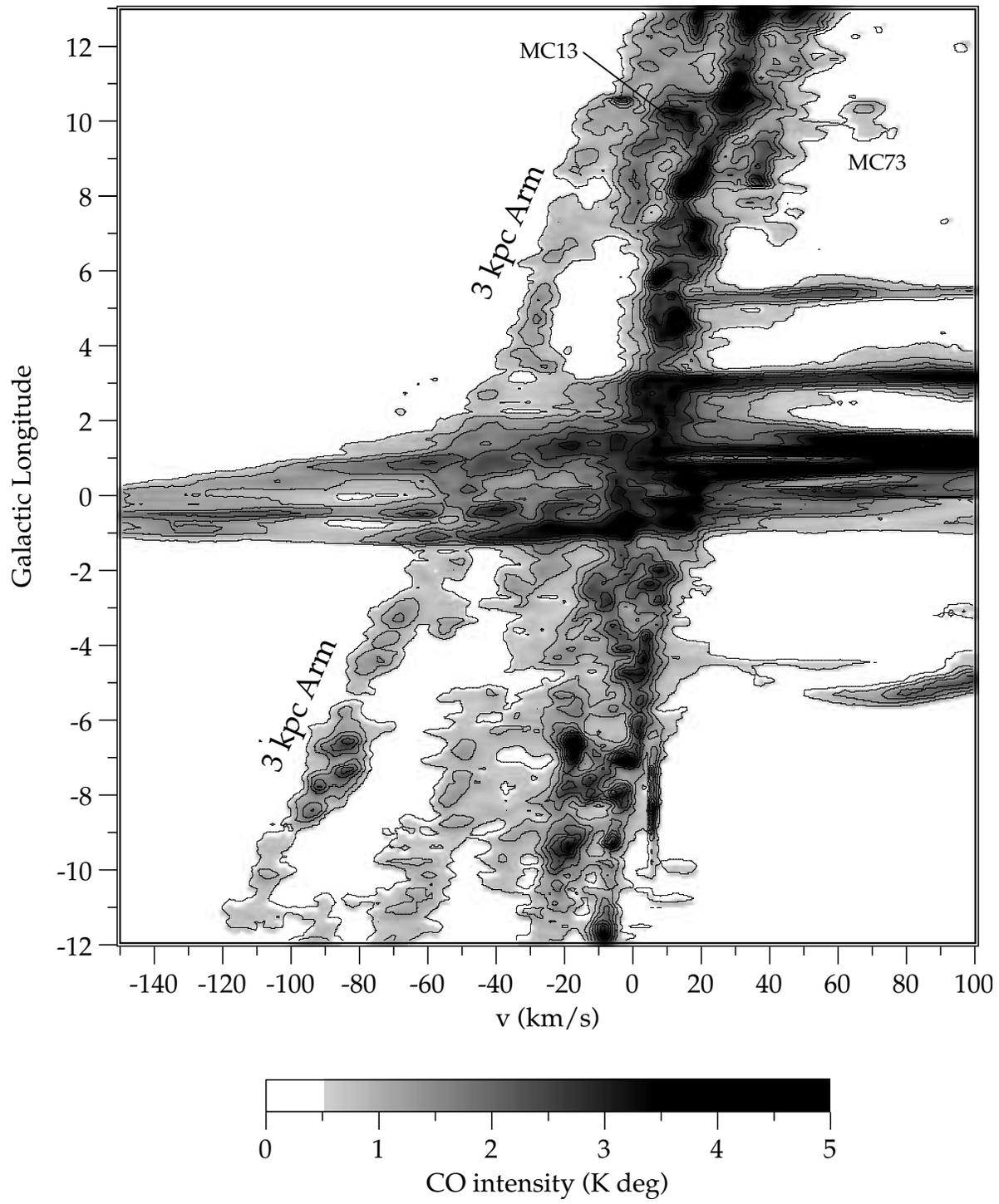
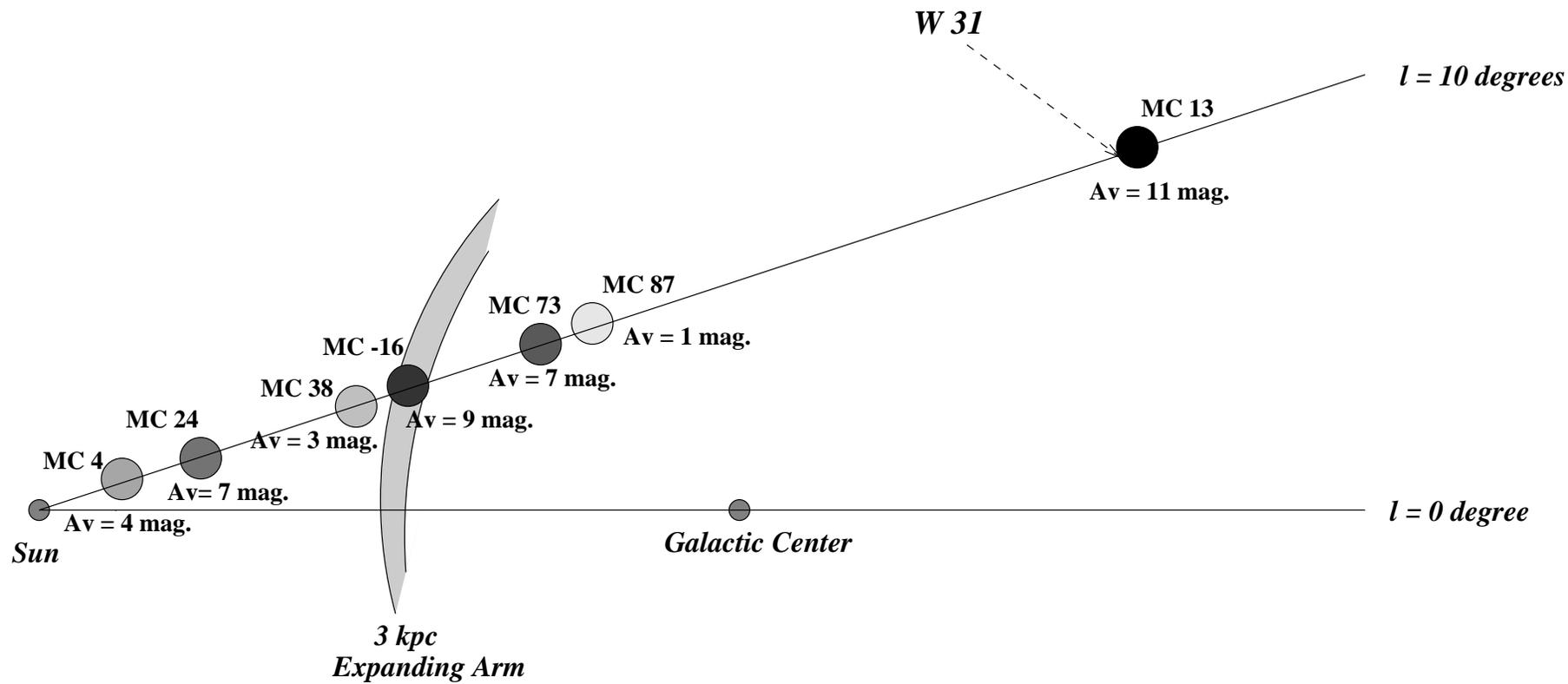
\end{document}